\documentclass[twocolumn,showpacs,preprintnumbers,amsmath,amssymb]{revtex4}
\usepackage{pifont}
\usepackage{tipa}
\usepackage{graphicx}
\usepackage{dcolumn}
\usepackage{bm}

\begin{document}

\title{Possible $2S$ and $1D$ charmed and charmed-strange mesons}
\author{Bing Chen\footnote{chenbing@shu.edu.cn}, Ling Yuan
and Ailin Zhang\footnote{corresponding author:
zhangal@staff.shu.edu.cn}} \affiliation{Department of Physics,
Shanghai University, Shanghai 200444, China}


\begin{abstract}
Possible $2S$ and $1D$ excited $D$ and $D_s$
states are studied, the charmed states $D(2550)^0$, $D^\ast(2600)$, $D(2750)^0$ and $D^\ast(2760)$
newly observed by the BaBar Collaboration are analyzed. The masses of these states
are explored within the Regge trajectory phenomenology, and the strong decay widths
are computed through the method proposed by Eichten {\it et al.}\cite{Eichten}. Both the mass and the decay width
indicate that $D(2550)^0$ is a good candidate of $2^1S_0$. $D^\ast(2600)$ and
$D_{s1}^\ast(2700)^\pm$ are very possible the admixtures of
$2^3S_1$ and $1^3D_1$ with $J^P=1^-$ and a mixing angle $\phi\approx 19^0$. $D^\ast(2760)$ and
$D_{sJ}^\ast(2860)^\pm$ are possible the $1^3D_3$ $D$ and $D_s$, respectively. $D(2750)^0$ and
$D^\ast(2760)$ seem two different states, and $D(2750)^0$ is very possible the
$1D(2^-, \frac{5}{2})$ though the possibility of $1D(2^-, \frac{3}{2})$ has not been excluded. There may exist an unobserved meson $D_{sJ}(2850)^\pm$ corresponding to $D^\ast_{sJ}(2860)^\pm$.
\end{abstract}


\maketitle

\section{Introduction}

The properties of $2S$ and $1D$ $Q\overline{q}$ mesons have been studied for a long time. However, no such
higher excited $Q\overline{q}$ state has been established for lack of
experimental data. In the past years, some higher excited
charmed or charmed-strange states were reported though most of them have not yet been pinned down~\cite{PDG}.
It will be useful to systemically study the possible $2S$ and $1D$ charmed and charmed-strange mesons in time.

The first possible charmed radial excitation, $D^{\ast\prime}(2640)$, was
reported by DELPHI~\cite{DELPHI}. This state is difficult to be understood
as a charmed radially excited state for the observed
decaying channel $D^{\ast+}\pi^+\pi^-$ and decay width $<15$ MeV~\cite{Pene}.
Its existence has not yet been confirmed by other collaboration. $D_{sJ}(2632)^+$ is another
puzzling state firstly observed by SELEX~\cite{SELEX}. It has not been observed by other collaboration either.
$D_{sJ}(2632)^+$ seems impossible a conventional $c\overline{s}$
meson for its narrow decay width and anomalous branching ratio
$\Gamma(D^0 K^+)/\Gamma(D_s^+ \eta)=0.14\pm0.06$~\cite{Barnes1} even if it does exist. In
an early analysis of the spectrum within Regge trajectories phenomenology~\cite{zhang1}, it is pointed
out that $D^{\ast\prime}(2640)$ and $D_{sJ}(2632)^+$ seem not the orbital excited tensor
states or the first radially excited state.

The observation of another three $D_s$ mesons:
$D_{s1}^\ast(2700)^\pm$~\cite{Belle1,Bar1,Bar2},
$D_{sJ}^\ast(2860)^\pm$~\cite{Bar1,Bar2} and
$D_{sJ}(3040)^+$~\cite{Bar2}, has evoked much more study of highly
excited $Q\overline{q}$ mesons. The masses and the decay widths of
$D_{s1}^\ast(2700)^\pm$ and $D_{sJ}^\ast(2860)^\pm$ were reported
by experiments. Furthermore, the ratios of branching fractions,
${\mathcal {B}(D_{s1}^\ast(2700)^\pm\to D^\ast K)\over \mathcal
{B}(D_{s1}^\ast(2700)^\pm\to DK)}=0.91\pm 0.13_{stat}\pm
0.12_{syst}$ and ${\mathcal {B}(D^\ast_{sJ}(2860)^+\to D^\ast
K)\over \mathcal {B}(D^\ast_{sJ}(2860)^+\to DK)}=1.10\pm
0.15_{stat}\pm 0.19_{syst}$, were measured. These states have been explored
within some models. $D_{s1}^\ast(2700)^\pm$ was identified
with the first radial excitation of
$D_s^\ast(2112)^\pm$~\cite{Close,ourpaper1}, or the
$D_s(1^3D_1)$~\cite{zhu}, or the mixture of them~\cite{Close}.
$D_{sJ}^\ast(2860)^\pm$ was interpreted as the
$D_s(2^3P_0)$~\cite{rupp,zhu} or the
$D_s(1^3D_1)$~\cite{colangelo,zhu,ourpaper1}. $D_{sJ}(3040)^+$ was
identified with the radially excited $D_s(2{1\over
2}^+)$~\cite{ourpaper1,liu0}. However, theoretical predictions of
these states are not completely consistent with experiments either
on their spectrum or on their decay widths.

Four new charmed states, $D(2550)^0$, $D^\ast(2600)^0$, $D(2750)^0$ and $D^\ast(2760)^0$
(including two isospin partners $D^\ast(2600)^+$ and
$D^\ast(2760)^+$) were recently observed by the BaBar collaboration~\cite{BaBar1}.
Some ratios of branching fractions of $D^\ast(2600)^0$ and $D(2750)^0$ were also measured.
In their report, analysis of the masses and helicity-angle distributions indicates
that $D(2550)^0$ and $D^\ast(2600)$ are possible the first radially excited $S-$wave
states $D(2^1S_0)$ and $D(2^3S_1)$, respectively, while other two charmed candidates
are possible the $1D$ orbitally excited states.

Theoretical analyses indicate that $D(2550)^0$ is a good
candidate of $2^1S_0$ though the predicted narrow width of $2^1S_0$ is inconsistent with the
observation~\cite{liu,zhong}. $D^\ast(2600)^0$ is interpreted as
a mixing state of $2^3S_1$ and $1^3D_1$~\cite{liu,zhong}. The calculation in Ref.~\cite{liu}
indicates that $D^\ast(2760)^0$ can be regarded as the orthogonal
partner of $D^\ast(2600)^0$ (or $1^3D_3$), but this
possibility (or $D^\ast(2760)$ is predominantly the $1^3D_1$) was excluded in Ref.~\cite{zhong}, where
$D^\ast(2760)$ is identified with the $1^3D_3$ state. In
Ref.~\cite{zhong}, the identification of $D(2750)^0$ and
$D^\ast(2760)$ with the same resonance with $J^P=3^-$ does not
favored.

Obviously, these $D$ and $D_s$ candidates have not yet been pinned down. In addition to some
theoretical deviations from experiments, some theoretical
predictions of their strong decays are different in different models. Systematical study of these
possible $2S$ and $1D$ states in more models is required. In this paper, the method proposed by Eichten \emph{et
al.}~\cite{Eichten} is employed to study the strong decay of the heavy-light mesons. We will label them
with the notaion $nL(J^P, j_q)$ in most cases, where $n$ is the radial quantum number,
$L$ is the orbital angular momentum, $J^P$ refers to the total
angular momentum and parity, $j_q$ is the total angular momentum of
the light degrees of freedom.

The paper is organized as follows. In Sec.II,
the spectrum of $2S$ and $1D$ $D_s$ and $D$ will be examined within the Regge
trajectory phenomenology. In Sec.III, two-body strong decay of these states will be
explored with EHQ's method. Finally, we present our conclusions and
discussions in Sec.IV.

\section{Mass spectrum in Regge trajectories}

Linearity of Regge trajectories (RTs) is an important observation in particle physics~\cite{chew}.
In the relativized quark model~\cite{GI}, the RTs for normal mesons are linear. For $Q\overline{q}$ mesons,
the approximately linear, parallel and equidistant RTs were obtained both in $(J,M^2)$ and in $(n_r,M^2)$
planes in the framework of a QCD-motivated relativistic quark
model~\cite{EFG}.

However, when RTs are reconstructed with the experimental data,
the linearity is always approximate. For orbitally excited states,
Tang and Norbury plotted many RTs of mesons and indicated that the
RTs are non-linear and intersecting~\cite{Tang}:
\begin{eqnarray}\label{eq1}
M^2=a J^2+b J+c,
\end{eqnarray}
where the coefficients $a, b, c$ are fixed by the experimental data, and $|a|\ll |b|$~\cite{Tang}. The
coefficients are usually different for different RTs.

For radially excited light $q\overline{q}$ mesons, Anisovich
\emph{et al.} systematically studied the trajectories on the planes $(n,M^2)$ in the mass
region up to $M<2400$ MeV~\cite{plot}. The RTs on $(n, M^2)$ plots behave as
\begin{eqnarray}\label{eq2}
M^2=M^2_0+(n-1)\mu^2,
\end{eqnarray}
where $M_0$ is the mass of the basic meson, $n$ is the radial quantum
number, and $\mu^2$ is the slope parameter of the trajectory.

Possible $1S$ and $2S$ $D$ and $D_s$ states are listed in Table~\ref{table-1},
, where $^\dag D^\prime_{s}(2635)$ is the predicted mass of $2S(1^-,
\frac{1}{2})$ $D_s$ meson. It is easy to notice that these candidates of $1S$ and $2S$
meet well with the trajectories on the $(n, M^2)$ plot according to Eq.~(\ref{eq2}).
The narrow charmed strange state $D_{sJ}(2632)^+$ is located around the mass region of $2S$ $D_s$.
However, the exotic relative branching ratio $\Gamma(D^0 K^+)/\Gamma(D_s^+
\eta)=0.14\pm0.06$ excludes its $2S(1^-,\frac{1}{2})$ possibility.
Therefore, we denotes the $2S(1^-, \frac{1}{2})$ $D_s$ meson with $^\dag D^\prime_{s}(2635)$.
As pointed in Ref.~\cite{ourpaper1}, the $2P$ candidate $D_{sJ}(3040)^+$ meets well with the trajectory on the $(n, M^2)$ plot.

\begin{table}
\renewcommand\arraystretch{1.2}
\begin{tabular}{c | c c | c c}
\hline\hline
States & $(0^-, \frac{1}{2})$ & \hspace {0.15cm} $(1^-, \frac{1}{2})$  & $(0^-, \frac{1}{2})$ &\hspace {0.12cm} $(1^-, \frac{1}{2})$\\
\hline
2S& $D(2550)^0$ & \hspace {0.1cm} $D^\ast_1(2600)$  & $^\dag D^\prime_{s}(2635)$ & \hspace {0.1cm} $D^\ast_{s1}(2700)^\pm$\\
1S& $D(1869)^\pm$ & \hspace {0.1cm} $D^\ast(2007)^0$  & $D_s(1968)^\pm$ &\hspace {0.1cm} $D^\ast_s(2112)^\pm$\\
\hline $\mu^2$(GeV$^2$) &  2.97 & 2.78  & 3.07& 2.88
\\ \hline\hline
\end{tabular}
\caption{$1S$ and $2S$ $D$ and $D_s$ mesons.} \label{table-1}
\end{table}

The measured masses of $D(2750)^0$, $D^\ast(2760)$ and
$D_{sJ}^\ast(2860)^\pm$ seem a little lower than most
theoretical predictions of the $1D$ states~\cite{GI,PE,EFG}.
In Fig.1, non-linear RTs of $D$ and $D_s$ states consisting of
$1^3S_1(1^-)$, $1^3P_2(2^+)$ and $1^3D_3(3^-)$ were reconstructed, where
the polynomial fits indicate $|a|\ll |b|$. In
a relativistic flux tube model, a ratio $b_{hl}/b_{ll}=2$ was obtained
at the lowest order~\cite{Olsson}, where $b_{hl}$ is the coefficient for
the heavy-light meson and $b_{ll}$ is the coefficient for the light-light meson
in Eq.~(\ref{eq1}). The $b_{ll}$ (about $0.70\sim1.60$) has been obtained in Ref.~\cite{Tang}.
the fitted $b_{hl}$ of $D$ and $D_s$ in Fig.1 is about $2.74$ and $3.03$, respectively.
Obviously, the fitted ratio is consistent with the theoretical prediction.

\begin{figure}[ht]
\begin{center}
\includegraphics[width=8cm,keepaspectratio]{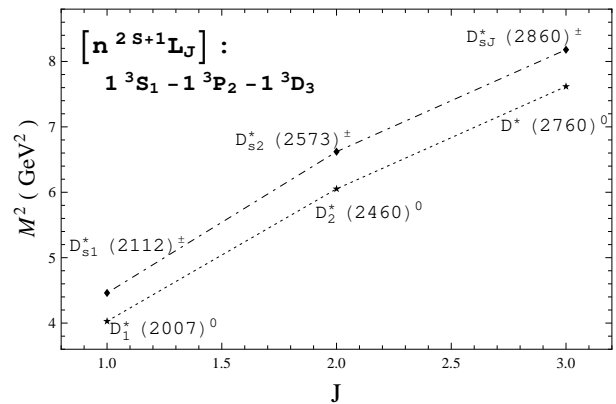}
\caption{Non-linear RTs of the $D$ and $D_s$ triplet with $N$, $S=1$. The
polynomial fits are $M^2=-0.23J^2+2.74J+1.53$ (GeV$^2$) and
$M^2=-0.29J^2+3.03J+1.72$ (GeV$^2$), respectively.}
\end{center}
\end{figure}

Through the analysis of the spectrum only, $D(2550)^0$, $D^\ast(2600)$ and
$D_{s1}^\ast(2700)^\pm$ are very possible the first radially excited
$D$ and $D_s$ states, and $D^\ast(2760)$ and
$D_{sJ}^\ast(2860)^\pm$ are possible the $1^3D_3$ states.

However, as well known, the RTs can only give a preliminary analysis of the observed states, the
investigation of the decay widths and the ratios of
branching fractions will be more useful to shed light on the underlying
properties of these states.

\section{Decay width in EHQ's formula}

The decay properties of heavy-light mesons have been studied in detail
in the heavy quark effective theory (HQET). Here, a concise method proposed by
Eichten {\it et al.}~\cite{Eichten} is employed to study the decays of $D$ and $D_s$ mesons.
As well known, in the heavy quark symmetry theory, the heavy-light
mesons degenerate in $j^P_l$, $i.e.$, two orbital ground states form a spin doublet $1S(0^-, 1^-)$
with $j^P_l=\frac{1}{2}^-$, and the decay amplitude satisfies certain symmetry
relations due to the heavy quark symmetry~\cite{Isgur1}. Therefore, the decays of the two mesons
in one doublet are governed by the same transition strength.

In the decay of an excited heavy-light meson $H$, characterized
by $nL(J^P, j_l)$, to a heavy-light meson $H^\prime$ ($n'L'({J'}^{P'}, {j'}_l)$)
and a light hadron $h$ with spin $s_h$ and orbital
angular momentum $l$ relative to $H'$, the two-body strong decay width (the EHQ's formula)
is written as~\cite{Eichten}
\begin{eqnarray}\label{eq3}
\Gamma^{H\rightarrow H^\prime h}=\zeta (\mathcal
{C}^{s_Q,j'_q,J'}_{j_h,j_q,J})^2\mathcal
{F}^{j_q,j'_q}_{j_h,l}(0)p^{2l+1}exp(-\frac{p^2}{6\beta^2}).
\end{eqnarray}
Where
\begin{eqnarray*}
\mathcal {C}^{s_Q,j'_q,J'}_{j_h,J,j_q}=\sqrt{(2J'+1)(2j_q+1)}\left\{
           \begin{array}{ccc}
                    s_Q  & j'_q & J'\\
                    j_h  & J    & j_q\\
                    \end{array}
     \right\}
\end{eqnarray*}
and $\overrightarrow{j}_h=\overrightarrow{s}_h+\overrightarrow{l}$.
$\mathcal {F}^{j_q,j'_q}_{j_h,l}(0)$ is the transition strength, and
$p$ is the momentum of decay products in the rest frame of $H$. The coefficients
$\mathcal {C}$ depend only upon the total angular momentum $j_h$ of
the light hadron, and not separately on its spin $s_h$ and the
orbital angular momentum $l$ of the decay. The flavor factor
$\zeta$ used in this paper for different decay channels can be found in
Ref.~\cite{PE}.

For lack of measurements of partial widths in the charmed states, the decay width
of $K$ mesons (i.e. $K_1(1270)\to\rho K$) was used to fix the transition strength
in Ref.~\cite{Eichten}. $c$ and $b$ quarks are much heavier than $u$, $d$ and $s$ quarks, so the
open charmed or bottomed mesons provide better place to test EHQ's formula.
Systematical study of $S-$ and $P-$wave heavy-light meons
($D$, $B$, $D_s$ and $B_s$ mesons) by EHQ's formula
have been presented in Ref.~\cite{ourpaper}. In the reference, the EHQ's formula is also obtained
by the well-known $^3P_0$ model~\cite{3P0}. In this way, the transition strength
$\mathcal {F}^{j_q,j'_q}_{j_h,l}(0)$ obtained in the $^3P_0$ model
includes only two parameters, the dimensionless parameter $\gamma$ and the wave
function inverse length scale $\beta$ ($0.35\sim0.40$ GeV)~\cite{ourpaper}.

The relevant transition strengths $\mathcal {F}^{j_q,j'_q}_{j_h,l}(0)$ used in this paper
are given in Table II. Some expressions in the table are from Ref.~\cite{Barnes, Isgur}, and others are obtained in the $^3P_0$ model in detail in Ref.~\cite{ourpaper}. For these transition strengths, a constant
\begin{eqnarray}\label{eq4}
\mathcal
{G}=\pi^{1/2}\gamma^2\frac{2^{10}}{3^4}\frac{\widetilde{M}_B\widetilde{M}_C}{\widetilde{M}_A}\frac{1}{\beta}
\end{eqnarray}
was omitted. Here the phase space normalization of Kokoski and Isgur is employed~\cite{GI,Isgur}.

In the analysis follows, the decay widths of possible $2S$ and $1D$ $D$
and $D_s$ states are computed in terms of Eq.~(\ref{eq3}).
\begin{table}
\renewcommand\arraystretch{1.6}
\begin{tabular}{ c | c | c }
\hline\hline
$nL(j^P_l)\rightarrow nL(j^P_l)+ \mathcal {P}$ & $\mathcal {F}^{j_q,j'_q}_{j_h,l}(0)$ & Polynomial of $p/\beta$\\
\hline
$2S(\frac{1}{2}^-)\rightarrow 1S(\frac{1}{2}^-)+ 0^-$ &  $\mathcal {F}^{\frac{1}{2},\frac{1}{2}}_{1,1}(0)$ & $\frac{5^2}{3^4}\frac{1}{\beta^2}(1-\frac{2}{15}\frac{p^2}{\beta^2})^2$  \\
$2S(\frac{1}{2}^-)\rightarrow 1P(\frac{1}{2}^+)+ 0^-$ &  $\mathcal {F}^{\frac{1}{2},\frac{1}{2}}_{0,0}(0)$ & $\frac{1}{2\cdot3^3}(1-\frac{7}{9}\frac{p^2}{\beta^2}+\frac{2}{27}\frac{p^4}{\beta^4})^2$   \\
$2S(\frac{1}{2}^-)\rightarrow 1P(\frac{3}{2}^+)+ 0^-$& $\mathcal {F}^{\frac{1}{2},\frac{3}{2}}_{2,2}(0)$   & $\frac{13^2}{3^7}\frac{1}{\beta^4}(1-\frac{2}{39}\frac{p^2}{\beta^2})^2$  \\
\hline
$1D(\frac{3}{2}^-)\rightarrow 1S(\frac{1}{2}^-)+ 0^-$& $\mathcal {F}^{\frac{3}{2},\frac{1}{2}}_{1,1}(0)$   & $\frac{5\cdot2}{3^4}\frac{1}{\beta^2}(1-\frac{2}{15}\frac{p^2}{\beta^2})^2$ \\
$1D(\frac{3}{2}^-)\rightarrow 1S(\frac{1}{2}^-)+ 1^-$& $\mathcal {F}^{\frac{3}{2},\frac{1}{2}}_{1,1}(0)$   & $\frac{2^2}{3^4}\frac{1}{\beta^2}(1-\frac{2}{15}\frac{p^2}{\beta^2})^2$\\
$1D(\frac{3}{2}^-)\rightarrow 1P(\frac{1}{2}^+)+ 0^-$& $\mathcal {F}^{\frac{3}{2},\frac{1}{2}}_{2,2}(0)$   & $\frac{5}{3^7}\frac{1}{\beta^4}(1+\frac{2}{15}\frac{p^2}{\beta^2})^2$ \\
$1D(\frac{3}{2}^-)\rightarrow 1P(\frac{3}{2}^+)+ 0^-$& $\mathcal {F}^{\frac{3}{2},\frac{3}{2}}_{0,0}(0)$   & $\frac{2^2\cdot5}{3^3}(1-\frac{5}{18}\frac{p^2}{\beta^2}+\frac{1}{135}\frac{p^4}{\beta^4})^2$ \\
& $\mathcal {F}^{\frac{3}{2},\frac{3}{2}}_{2,2}(0)$                                                        & $\frac{13^2}{3^7\cdot5}\frac{1}{\beta^4}(1-\frac{2}{39}\frac{p^2}{\beta^2})^2$ \\
\hline
$1D(\frac{5}{2}^-)\rightarrow 1S(\frac{1}{2}^-)+ 0^-$& $\mathcal {F}^{\frac{5}{2},\frac{1}{2}}_{3,3}(0)$   & $\frac{2^3}{3^6\cdot5}\frac{1}{\beta^6}$ \\
$1D(\frac{5}{2}^-)\rightarrow 1S(\frac{1}{2}^-)+ 1^-$& $\mathcal {F}^{\frac{5}{2},\frac{1}{2}}_{3,3}(0)$   & $\frac{2^5}{3^7\cdot5}\frac{1}{\beta^6}$ \\
                                                     & $\mathcal {F}^{\frac{5}{2},\frac{1}{2}}_{2,1}(0)$   & $\frac{2^4}{3^4}\frac{1}{\beta^2}(1-\frac{2}{15}\frac{p^2}{\beta^2})^2$ \\
$1D(\frac{5}{2}^-)\rightarrow 1P(\frac{1}{2}^+)+ 0^-$& $\mathcal {F}^{\frac{5}{2},\frac{1}{2}}_{2,2}(0)$   & $\frac{2^2\cdot5}{3^7}\frac{1}{\beta^4}(1-\frac{1}{15}\frac{p^2}{\beta^2})^2$ \\
$1D(\frac{5}{2}^-)\rightarrow 1P(\frac{3}{2}^+)+ 0^-$& $\mathcal {F}^{\frac{5}{2},\frac{3}{2}}_{2,2}(0)$   & $\frac{2^5\cdot7}{3^7\cdot5}\frac{1}{\beta^4}(1-\frac{1}{42}\frac{p^2}{\beta^2})^2$ \\
                                                     &$\mathcal {F}^{\frac{5}{2},\frac{3}{2}}_{4,4}(0)$   & $\frac{2^4}{3^8\cdot5\cdot7}\frac{1}{\beta^8}$ \\
\hline\hline
\end{tabular}
\caption{The transition strength $\mathcal
{F}^{j_q,j'_q}_{j_h,l}(0)$, where the sign ``$\mathcal {P}$"
denotes a light pseudoscalar meson or a light vector meson.}
\label{table-2}
\end{table}

\ding {172}.\hspace {0.2cm} $2^1S_0$ or [$2S(0^-, \frac{1}{2})$]

$D(2550)^0$ observed in the decay channel $D^{\ast +}\pi^-$ is a good
candidate of $2^1S_0$ charmed meson. Following the procedure in Ref.~\cite{ourpaper},
we take the decay width of $D^\ast_2(2460)^0$ as an input and obtain the
$d-$wave transition strength $\mathcal {F}^{\frac{3}{2},\frac{1}{2}}_{2,2}(0)=0.964$ GeV$^{-4}$
with $\beta=0.38$ GeV$^{-1}$, where
\begin{eqnarray*}
\mathcal {F}^{\frac{3}{2},\frac{1}{2}}_{2,2}(0)=\mathcal
{G}\frac{2^2}{3^4}\frac{1}{\beta^4}.
\end{eqnarray*}

All other transition strength $\mathcal
{F}^{j_q,j'_q}_{j_h,l}(0)$ in Table~\ref{table-2} could be fixed once
the mock-meson masses $\widetilde{M}_i$ effect has been taken into account.
According to our computation~\cite{ourpaper}, the total decay width of $D(2550)^0$
$\Gamma=124.1$ MeV. The dominating decay mode is the $D^\ast\pi$ channel with $\Gamma(D^\ast\pi)=121.0$ MeV,
and the decay width of another allowed $D^\ast_0(2400)\pi$ channel is $3.1$ MeV
(the mass of $D^\ast_0(2400)$ is taken as $2318$ MeV~\cite{PDG}).

These results agree well with the experiments. It explains the fact that $D(2550)^0$ was first observed in
$D^{\ast +}\pi^-$~\cite{BaBar1}. In Fig. 2, the variation of the decay width with $\beta$ is plotted. Obviously, the observed decay width of $D(2550)^0$ is well obtained in the reasonable region of $\beta$
($0.35\sim 0.40$ GeV),

In $D_s$ states, the mass of the $2^1S_0$ state is predicted around $2635\pm 20$ MeV (a little smaller
than the threshold of $D^\ast\eta$ and $D^\ast_0(2400) K$), and $D^\ast K$ is the only two-body strong decay channel. Our result for this decay channel is $\Gamma(D^\ast K)\approx 82.2\pm 15.1$ MeV, so the observed $D_{sJ}(2632)^+$ is impossible the $2^1S_0$ $D_s$ meson.

\begin{figure}[ht]
\begin{center}
\includegraphics[width=8cm,keepaspectratio]{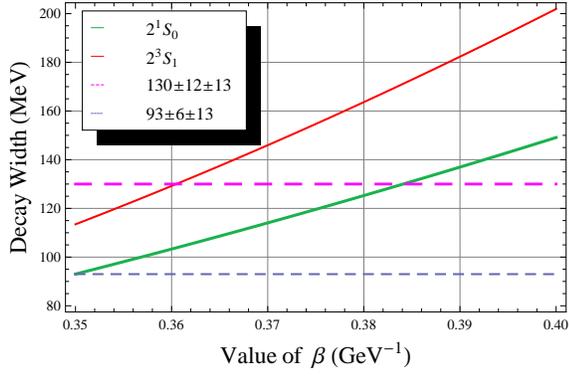}
\caption{The decay width with $\beta$, where one
takes $D(2550)^0$ as a pure $2^1S_0$ (green line) state and
$D^\ast(2600)^0$ as a pure $2^3S_1$ (red line) state. Dash lines
refer to central values of decay width given by experiment.}
\end{center}
\end{figure}

\ding {173}.\hspace {0.2cm} Mixing states of $2^3S_1$ and $1^3D_1$

The spectrum and the helicity-angle distributions support
the suggestion that $D^\ast(2600)$ is the $2^3S_1$. However, if $D^\ast(2600)$ is
a pure $2^3S_1$ state, its decay width is about $163.7$ MeV
($\beta=0.38$ GeV$^{-1}$). This decay width seems broader (see
Fig. 2) than the experiment ($93\pm6\pm13$ MeV). Similarly,
the decay width of $D_{s1}^\ast(2700)^\pm$ is $230.5$ MeV if
it is a pure $2^3S_1$, which deviates also from the
experiment ($125\pm 30$ MeV).

In charmonium system, $\psi(2S)$ and
$\psi(3770)$ are two orthogonal partners of mixtures of $2^3S_1$ and $1^3D_1$
with $J^{PC}=1^{--}$~\cite{EG}. This mixing scheme has also been employed to
explain the decay width and the ratio of branching fractions of
$D_{s1}^\ast(2700)^\pm$ and $D^\ast_{sJ}(2860)^\pm$~\cite{Close}. Similarly,
we denote two orthogonal partners ($J^P=1^-$) of $D$ and $D_s$ as
\begin{equation}
\begin{aligned}
|(SD)_1\rangle_L = cos\phi |2^3S_1\rangle - sin\phi|1^3D_1\rangle,\\
|(SD)_1\rangle_R = sin\phi |2^3S_1\rangle + cos\phi |1^3D_1\rangle.
\end{aligned}
\end{equation}

When $D^\ast(2600)$ and $D_{s1}^\ast(2700)^\pm$ are identified with the $|(SD)_1\rangle_L$ of $D$
and $D_s$, respectively, their decay widths variation with the mixing angle $\phi$ are calculated
and presented in Fig. 3. The experimental decay widths of
$D^\ast(2600)$ and $D_{s1}^\ast(2700)^\pm$ are well obtained at $\phi\approx 20^0$,

\begin{figure}[ht]
\begin{center}
\includegraphics[width=8cm,keepaspectratio]{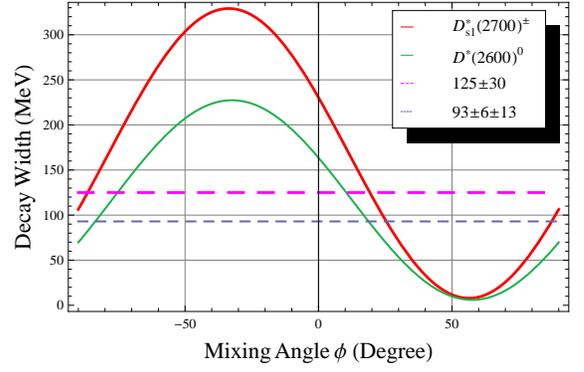}
\caption{Decay widths of $D^\ast(2600)^0$ (green
line) and $D_{s1}^\ast(2700)^\pm$ (red line) with the mixing angle
$\phi$. Dash lines refer to central values of decay width given by
experiment.}
\end{center}
\end{figure}

The ratios of branching fractions variation with the mixing
angle $\phi$ are presented in Fig. 4. When $\phi\approx 20^0$, the observed ratio of
branching fraction $\mathcal {B}(D^\ast(2600)^0\rightarrow
D^+\pi^-)/\mathcal {B}(D^\ast(2600)^0\rightarrow D^{\ast+}\pi^-)=0.32\pm 0.02\pm 0.09$
is also obtained. However, theoretical $\mathcal
{B}(D_{s1}^\ast(2700)\rightarrow DK)/\mathcal
{B}(D_{s1}^\ast(2700)\rightarrow D^\ast K)$ seems a little
smaller than the experimental data. More experimental
measurements of the ratios are required.

\begin{figure}[ht]
\begin{center}
\includegraphics[width=8cm,keepaspectratio]{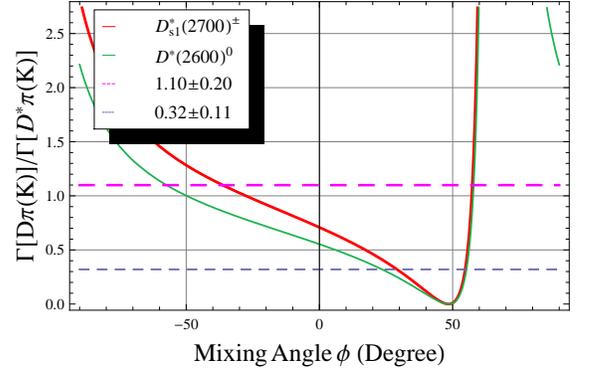}
\caption{Ratios of branching fractions with the mixing
angle $\phi$ of $D^\ast(2600)^0$ (green line) and $D_{s1}^\ast(2700)^\pm$ (red line).
Dash lines refer to central values of experiments.}
\end{center}
\end{figure}

The partial widths of all two-body strong decay modes of
$D^\ast(2600)^0$ and $D_{s1}^\ast(2700)^\pm$ are presented in Table III. Their
total decay widths are in accord with experiments. In summary, $D^\ast(2600)^0$ and
$D_{s1}^\ast(2700)^\pm$ are very possible the $|(SD)_1\rangle_L$ of $D$ and $D_s$, respectively.

The decay channels $D^\ast(2760)^0\to D^+\pi^-$ and $D_{sJ}^\ast(2860)^+\to D^0K^+$ have been observed.
However, it is difficult to identify $D^\ast(2760)^0$ and $D_{sJ}^\ast(2860)^+$ with
the $|(SD)_1\rangle_R$, the orthogonal partners of $D^\ast(2600)$ and $D_{s1}^\ast(2700)^\pm$, respectively.
In that case, the decay width of $D_{sJ}^\ast(2860)^+$ is broader than $200$ MeV and the decay width of $D^\ast(2760)^0$) is broader than $110$ MeV, these predictions are much broader than the experimental results.

\begin{table}
\renewcommand\arraystretch{1.2}
\begin{tabular}{cc | cccc}
\hline\hline
Modes$^{(1)}$ & $\Gamma_i$(MeV) & Modes$^{(2)}$ &  $\Gamma_i$ (MeV) &  Modes$^{(2)}$ & $\Gamma_i$(MeV)\\
\hline
$D^\ast$ $K$ & 76.8  & $D^\ast$ $\pi$ & 60.7 & $D_s$ $K$ & 3.2  \\
$D$ $K$       &   36.7      & $D$ $\pi$ & 22.5 & $D'_1(2430)\pi$ & 2.2 \\
$D^\ast_s$ $\eta$  & 4.6 & $D^\ast$ $\eta$ & 1.2 & $D_1(2420)\pi$ & 0.1 \\
$D_s$ $\eta$ & 8.2 & $D$ $\eta$ & 2.0 & $D^\ast_2(2460)\pi$& 0 \\
\hline
$\Gamma^{(1)}_{total}$ & 126.2 &   &    & $\Gamma^{(2)}_{total}$ & 92.0 \\
Expt. & $125\pm30$ &   &    & Expt. & $93\pm19$ \\
\hline\hline
\end{tabular}
\caption{Two-body strong decays of the admixture of $2^3S_1$ and
$1^3D_1$ with $J^P=1^-$. Here $\beta=0.38$ GeV$^{-1}$ and mixing
angle $\phi\approx 19^0$.
 ``Modes$^{(1)}$" refers to decay modes of $D_{s1}^\ast(2700)$ and
``Modes$^{(2)}$" refers to decay modes of $D^\ast(2600)$.}
\label{table-3}
\end{table}

\ding {174}.\hspace {0.2cm}$1^3D_3$ or [$1D(3^-, \frac{5}{2})$]

$D^\ast(2760)$ and $D_{sJ}^\ast(2860)^\pm$ are very possible the $1^3D_3$ $D$ and $D_s$, respectively.

$D^\ast(2760)^0$ was observed in the decay channel $D^+\pi^-$
and was suggested to be a $D-$wave charmed meson~\cite{BaBar1}.
If $D^\ast(2760)^0$ has the same $J^P$ with the $1^3D_1$, it would have a broad width
through the mixing scheme mentioned above.

Under the assumption that both $D^\ast(2760)$ and $D_{sJ}^\ast(2860)^\pm$ are the $1^3D_3$
states, their partial widths and total decay widths are given in Table IV. The predicted decay widths
of them are in accord with experimental results.

\begin{table}
\renewcommand\arraystretch{1.2}
\begin{tabular}{cc|cccc}
\hline\hline
Modes$^{(a)}$ & $\Gamma_i$(MeV) & Modes$^{(b)}$ &  $\Gamma_i$(MeV) &  Modes$^{(b)}$ & $\Gamma_i$(MeV)\\
\hline
$D^\ast$ $K$ & 12.3  & $D^\ast$ $\pi$ & 12.4 & $D_s$ $K$ & 0.9  \\
$D$ $K$       &   28.4      & $D$ $\pi$ & 22.0 & $D^\ast_s$ $K$ & 0.1 \\
$D^\ast_s$ $\eta$  & 0.6 & $D^\ast$ $\eta$ & 0.2 & $D'_1(2430)\pi$ & 1.1 \\
$D_s$ $\eta$ & 3.0 & $D$ $\eta$ & 0.8 & $D_1(2420)\pi$ & 0.4  \\
$D$ $K^\ast$  & 0.5 & $D$ $\rho$ & 0.1 & $D^\ast_2(2460)\pi$& 1.3 \\
$D_s$ $\omega$ & 0.2 & $D$ $\omega$ & 0 & - & - \\
\hline
$\Gamma^{(a)}_{total}$ & 44.9 &   &    & $\Gamma^{(b)}_{total}$ & 39.3 \\
Expt. & $48\pm7$ &   &    & Expt. & $60.9\pm8.7$ \\
\hline\hline
\end{tabular}
\caption{Two-body strong decays of the states $1^3D_3$.
``Modes$^{(a)}$" refers to decay modes of $D_{sJ}^\ast(2860)$ and
``Mode$^{(b)}$" refers to those of $D^\ast(2760)$.} \label{table-4}
\end{table}

$D(2750)^0$ has mass close to that of $D^\ast(2760)$, if these two states are the same state of $1^3D_3$,
the predicted ratio $\Gamma(D^\ast(2760)\to D\pi)/\Gamma(D^\ast(2760)\to D^\ast\pi)=1.78$ (see Table IV)
is much larger than the observed $\mathcal {B}(D^\ast(2760)^0\rightarrow D^+\pi^-)/\mathcal
{B}(D(2750)^0\rightarrow D^{\ast +}\pi^-)=0.42\pm 0.05\pm 0.11$. This fact supports the suggestion
that $D(2750)^0$ and $D^\ast(2760)$ are two different charmed states~\cite{BaBar1,zhong}.

For $D_{sJ}^\ast(2860)^\pm$, the predicted $\Gamma(D_{sJ}^\ast(2860)^\pm\to D^\ast
K)/\Gamma(D_{sJ}^\ast(2860)^\pm\to DK)=0.43$ is much smaller than the experimental
${\mathcal {B}(D^\ast_{sJ}(2860)^+\to D^\ast K)\over \mathcal {B}(D^\ast_{sJ}(2860)^+\to DK)}=1.10\pm
0.15_{stat}\pm 0.19_{syst}$.

It is noticed that the mass gaps of the corresponding
ground state between $D$ and $D_s$ are about $100$ MeV~\cite{PDG}. The mass gap between
$D_{s1}^\ast(2700)^\pm$ and $D^\ast(2600)$, and the mass gap between
$D_{sJ}^\ast(2860)^\pm$ and $D^\ast(2760)$ are also about $100$ MeV.
The mass gap supports also the suggestion that $D_{s1}^\ast(2700)^\pm$ is a similar state as
$D^\ast(2600)$ with the same $J^P$. Therefore, there should exist a charmed-strange $D_{sJ}(2850)^\pm$ which has the same $(J^P, j_q)$ of $D(2750)^0$ with mass close to $D_{sJ}^\ast(2860)^\pm$.

\ding {175}.\hspace {0.2cm} $1D(2^-, \frac{3}{2})$ and $1D(2^-,
\frac{5}{2})$

$D(2750)^0$ was observed in $D^{\ast +}\pi^-$ and is possible a $1D(2^-, \frac{3}{2})$ or $1D(2^-, \frac{5}{2})$, there exists similar assignment for the suggested $D_{sJ}(2850)^\pm$.
The partial widths of some two-body decay modes of $D(2750)^0$ and $D_{sJ}(2850)^\pm$ in the two possible assignments have been computed and presented in Table V.

\begin{table}
\renewcommand\arraystretch{1.2}
\begin{tabular}{ccc|ccc}
\hline\hline
Modes$^\dag$ & $(2^-, \frac{3}{2})$ & $(2^-, \frac{5}{2})$ &  Modes$^\ddag$ &  $(2^-, \frac{3}{2})$ & $(2^-, \frac{5}{2})$\\
\hline
$D^\ast$ $\pi$      & 58.9  & 20.0 & $D^\ast$ $K$      & 96.2 & 19.2  \\
$D^\ast$ $\eta$     & 5.4   & 0.2  & $D^\ast_s$ $\eta$ & 21.7 & 0.9 \\
$D^\ast_s$ $K$      & 8.6   & 0.2  & $D$ $K^\ast$      & 4.3  & 18.0 \\
$D$ $\rho$           & 1.9   & 9.2  & - & - & -  \\
$D$ $\omega$         & 0.7   & 3.3  & $D_s$ $\omega$     & 2.7  & 13.3 \\
$D^\ast_0(2400)\pi$ & 0.6   & 10.9 & $D^\ast_0(2400) K$& 0.2  & 0.2 \\
$D'_1(2430)\pi$      & 0.2   & 1.4  & - & - & - \\
$D_1(2420)\pi$       & 0.5   & 1.4  & - & - & - \\
$D^\ast_2(2460)\pi$ & 1.2   & 0.3  & - & - & - \\
\hline
$\Gamma^{(\dag)}_{total}$ (MeV) & 77.9 &  47.9 & $\Gamma^{(\ddag)}_{total}$ (MeV)  & 125.1 & 51.6 \\
Expt.\hspace {0.72cm}  & - & $71\pm17$ &  Expt.\hspace {0.72cm}  & - & - \\
\hline\hline
\end{tabular}
\caption{Two-body strong decays of the states $(2^-, \frac{3}{2})$
and $(2^-, \frac{5}{2})$. ``Modes$^\dag$" and ``Modes$^\ddag$" refer
to decay modes of $D(2750)^0$ and $D_{sJ}(2850)$, respectively.}
\label{table-4}
\end{table}

If $D_{sJ}(2850)^\pm$ is the $1D(2^-, \frac{3}{2})$, the predicted ratio of branching
fraction $\mathcal {B}(D_{sJ}(2850)\rightarrow D^\ast K)/\mathcal
{B}(D_{sJ}(2860)\rightarrow D K)$ is about $2.42$. Theoretical predictions of the decay width and the ratio of branching fraction $\mathcal
{B}(D^\ast(2760)^0\rightarrow D^+\pi^-)/\mathcal
{B}(D(2750)^0\rightarrow D^{\star+}\pi^-=0.52$ of $D(2750)^0$ are in accord with experiment.

If $D(2750)^0$ and $D_{sJ}(2850)^\pm$ are
the $1D(2^-, \frac{5}{2})$, $D(2750)^0$, $D^\ast(2760)^0$ and
$D_{sJ}(2850)^\pm$, $D_{sJ}^\ast(2860)^\pm$ form the $1D(2^-, 3^-)$ doublet of $D$ and $D_s$,
respectively.
For charmed mesons $D(2750)^0$ and $D^\ast(2760)^0$, we obtained
$\mathcal {B}(D^0[\frac{5}{2}^-]\rightarrow D^+\pi^-)/\mathcal
{B}(D^0[\frac{5}{2}^-]\rightarrow D^{\star +}\pi^-)\approx 0.44$, which is in accord with the observed
$\mathcal {B}(D^\ast(2760)^0\rightarrow D^+\pi^-)/\mathcal
{B}(D(2750)^0\rightarrow D^{\ast +}\pi^-)=0.42\pm 0.05\pm 0.11$.
We obtained $\mathcal {B}(D^+_{sJ}[\frac{5}{2}^-]\rightarrow D^\ast K)/\mathcal
{B}(D^+_{sJ}[\frac{5}{2}^-]\rightarrow DK)\approx 1.71$ for
the charmed-strange mesons $D_{sJ}(2850)^\pm$ and
$D_{sJ}^\ast(2860)^\pm$, and the observed ${\mathcal {B}(D^\ast_{sJ}(2860)^+\to D^\ast K)\over \mathcal {B}(D^\ast_{sJ}(2860)^+\to DK)}=1.10\pm
0.15_{stat}\pm 0.19_{syst}$. Theoretical
predictions are in accord with experiments within the
uncertainties of the $^3P_0$ model.

In our computation, a simple device known as the \emph{Shmushkevich}
factory~\cite{HQET} is employed. In this device,
when a sample of a doublet $1D(2^-, 3^-)$ with random polarizations
of the $c$ quark are considered, $5/12$ of this sample is $1D(2^-,
\frac{5}{2})$, and $7/12$ is $1D(3^-, \frac{5}{2})$. In the
transition with one final ground ($s-$wave) charmed state, the
charmed state has also randomly polarized $c$ quark, which means
that $1/4$ of the charmed state is $D$ and $3/4$ of them is
$D^\ast$. Therefore, in the decays of $1D(2^-, 3^-)$ into ground
charmed states, $3/7$ of $1D(3^-, \frac{5}{2})$ decay into $D$ and
$4/7$ of $1D(3^-, \frac{5}{2})$ decay into $D^\ast$. In this case,
the partial width of $D^\ast K$ observed
by experiment is the total one of both $D(2750)^0$ and
$D^\ast(2760)^0$.

Of course, the two states in the doublet $1D(2^-, 3^-)$ (
$1D(2^-, \frac{5}{2})$ and $1D(3^-, \frac{5}{2})$) have masses close to each
other while their mass splitting is comparable to the uncertainty of
their masses, it will be difficult to distinguish these two states through the channel of $D \pi$ and $D^\ast
\pi$. However, the state $1D(2^-, \frac{5}{2})$ decays
through the $P-$wave and the $F-$wave while the state $1D(3^-, \frac{5}{2})$
can only decay through the $F-$wave. Therefore, the widths of decay channels
$D \rho$ and $D \omega$ of $D(2750)^0$ would much broader than those of
$D^\ast(2760)^0$. The observation of the
channels $D \rho$ and $D \omega$ in forthcoming experiments will be useful to pin down these states.

\section{Conclusions and discussions}

In this work, we study the possible $2S$ and $1D$ $D$ and $D_s$ states, especially the
four new $D$ candidates observed by the BaBar Collaboration. Both the mass and the decay width indicate that $D(2550)^0$ is a good candidate of the $2^1S_0$ charmed state.
The observed $D_{sJ}(2632)^+$ is impossible the $2^1S_0$ $Ds$ meson, which is predicted to have mass about $2635\pm20$ MeV and decay
width about $82.2\pm15.1$ MeV.

$D^\ast(2600)$ and
$D_{s1}^\ast(2700)^\pm$ are very possible the admixtures of
$2^3S_1$ and $1^3D_1$ with $J^P=1^-$ and a mixing angle $\phi\approx 19^0$. Our analysis does not support the possibility that $D^\ast(2760)$ and $D_{sJ}^\ast(2860)^\pm$ are the orthogonal partners of
$D^\ast(2600)$ and $D_{s1}^\ast(2700)^\pm$, respectively.

An unobserved meson, corresponding to $D(2750)^0$, $D_{sJ}(2850)^\pm$, may exist, more measurement of $D_{sJ}^\ast(2860)^\pm$ is required.
$D^\ast(2760)$ and $D_{sJ}^\ast(2860)^\pm$ could be identified with the $1^3D_3$ $D$ and $D_s$ states, respectively.
$D(2750)^0$ and $D^\ast(2760)$ favor to form the doublet $1D(2^-,
3^-)$. The possibility that $D(2750)^0$ is the  $1D(2^-, \frac{3}{2})$ state has not been excluded, so the observation of the channels $D \rho$ and $D \omega$ in experiment would be important for the identification of $D(2750)^0$ and
$D^\ast(2760)^0$. Based on the simple device known as the \emph{Shmushkevich} factory, some ratios of branching fractions given by experiments are well understood.

\section*{Acknowledgements}
This work is supported by the National Natural Science Foundation of
China (NSFC) under Grants No. 10775093 and No. 11075102. It is also
supported by the Graduates Innovation Fund of Shanghai University:
SHUCX092016.

\end{document}